\documentclass[12pt,epsfig]{article}
\usepackage{graphicx}

\input{tcilatex}

\begin{document}

\baselineskip=18.6pt plus 0.2pt minus 0.1pt


\begin{titlepage}
\title{
\begin{flushright}
 {\normalsize \small
GNPHE/09011}
 \\[0.2cm]
 \mbox{}
\end{flushright}
{\bf On    F-theory Quiver Models        }\\[.2cm]
{\bf  and }\\[.2cm]
{\bf  Kac-Moody Algebras}
\author{ Rachid Ahl Laamara$^1$\thanks{\tt{doctorants.lphe@fsr.ac.ma}}, Adil Belhaj$^{1,2}$\thanks{\tt{belhaj@unizar.es}},
 Luis J. Boya$^3$\thanks{\tt{luisjo@unizar.es}}, Leila Medari$^4$\thanks{\tt{l.medari@ucam.ac.ma}} ,  Antonio Segui$^3$\thanks{\tt{segui@unizar.es}}\\[.3cm]
{\small  $^1$UFR- Lab de Physique des Hautes Energies and Groupement National de PHE}
\\[-6pt] {\small GNPHE, Si\`{e}ge focal: Facult\'{e} des
Sciences, Rabat,  Morocco }\\
{ \small $^2$ Centre National de l'Energie, des Sciences et des Techniques Nucl\'eaires,   Rabat, Morocco} \\
{\small $^3$ Departamento de F\'isica Te\'orica, Universidad de
Zaragoza, E-50009-Zaragoza, Spain}
\\
{\small$^4$  LPHEA, Physics Department, Faculty of Science Semlalia,
Marrakesh, Morocco} } } \maketitle \thispagestyle{empty}
\begin{abstract}
We   discuss  quiver  gauge models  with bi-fundamental and
 fundamental matter obtained  from  F-theory compactified on ALE spaces over a
 four dimensional base space.  We focus  on the
base geometry which  consists of  intersecting $F_0={\bf CP^1}\times
{\bf CP^1}$ Hirzebruch  complex surfaces arranged as Dynkin graphs classified by three kinds of 
  Kac-Moody (KM) algebras: ordinary, i.e finite dimensional, affine and
indefinite, in particular hyperbolic. We  interpret the  equations defining these three
classes of generalized Lie algebras  as the anomaly cancelation condition of the
corresponding $N =1$ F-theory quivers in four dimensions.  We analyze  in some detail hyperbolic
geometries obtained from  the affine $\widehat A$ base geometry by adding a node, and
we find  that it can be used to  incorporate fundamental fields to a product  of $ SU$-type gauge groups and fields.
\end{abstract}
{\tt  KEYWORDS}: F-theory, quiver  gauge  theories, Kac-Moody
algebras.
\end{titlepage}
\newpage

\section{Introduction}

Four dimensional gauge theories with only four supercharges ($4D,\; N=1$), obtained from
local F-theory models have been originated by Vafa, and are   very interesting in 
connections  with Grand Unification
Theories (GUT), the Standard  Model (SM) and the  Large Hadron Collider Physics(LHC)\cite{DW1}-\cite{HV1}. In the study of F-theory GUT models, 
it  involves  considerations of geometric
singularities which are related to seven-branes extant in  type IIB superstring theory,  and
are localized in two dimensional transverse space. It has been shown that
the gauge fields similar to  those Minimal Supersymmetric Standard Model   (MSSM)  can be gotten  from the eight-dimensional
worldvolume of the  seven-brane wrapping del Pezzo surfaces with a GUT gauge
group. The latter can be broken to $SU(3)_{C}\times SU(2)_{L}\times U(1)_{Y}$
via an internal flux through the seven-brane in the $U(1)_{Y}$ direction of
the GUT group. Further  details on building of F-theory models can be found in 
\cite{FI}-\cite{BBS}. 

Motivated by these large activities, we have recently manipulated 
 some quivers with only bi-fundamental matter from F-theory
compactified on Asymptotically Locally   Euclidean (ALE) spaces over a four-dimensional base space \cite{ABBS}.
The  geometry of the base  has been obtained by blowing up the extended $ADE$-type  singularities of eight dimensional hyper-K%
\"{a}hler manifolds,  considered as sigma
model target spaces with eight supercharges \cite{BS}.  We have   shown that
the base can be identified with intersecting $F_{0}=\mathbf{CP^{1}}\times 
\mathbf{CP^{1}}$ according to the affine $ADE$ Dynkin diagrams required by
the anomaly cancelation condition. The corresponding seven-brane
configurations lead to a $\prod\limits_{i}SU(s_{i}N)$ gauge group with only
bi-fundamental chiral matter transforming in $(s_{i}N,s_{j}N)$
representations,  where $N$ is an unspecified positive integer. Besides, the
positive integers $s_{i}$ appearing in the gauge group are the Coxeter
labels (sometimes referred to as marks or Dynkin numbers). They form a
positive-definite integer vector $(s_{i})$ satisfying $\sum%
\limits_{i}K_{ij}s_{i}=0$ where $K$ is the Cartan matrix of the affine Lie
algebras.

The aim of this paper  is to discuss some seven-brane configurations leading
to F-theory quivers involving both bi-fundamental and fundamental matter.
In particular, the main result is the building of a base geometry in 
F-theory compactification producing such quivers. More precisely, we
consider different geometries which consist of intersecting $F_{0}=\mathbf{%
CP^{1}}\times \mathbf{CP^{1}}$ according to Dynkin diagrams of the three types of 
Kac-Moody (KM) algebras,  ordinary finite dimensional, affine (with Cartan matrix determinant$=0$)
 and indefinite, in particular ($det < 0$)\cite{Kac,GO,ABS}. The algebraic equations defining these Lie
algebras can  be interpreted as the anomaly cancellation condition of $N=1$
F-theory quivers in four dimenions. For a particular example of hyperbolic geometries obtained
from the $\widehat{A}$ geometry, we implement fundamental fields to a
linear chain of gauge groups $\prod\limits_{i}SU$ with bi-fundamental
matter. But,  first  we discuss the geometric engineering procedure to obtain  
bi-fundamental matter in F-theory quivers. We then extend this analysis  to geometric
engineering of fundamental matter in F-theory compactification, with 
particular emphasis on the two-dimensional complex base geometry classified
by KM algebras.

\section{ Bi-fundamental Matter from Affine Lie Algebras}

In this section, we will discuss how  the bi-fundamental matter can be
obtained from the intersecting seven-branes in F-theory compactifications \cite{ABBS}. We will start with a general construction of the  base of local F-theory  geometries.
The base  is obtained by resolving the so-called extended $ADE$ singularities of eight dimensional 
 hyper-K\"{a}hler manifolds considered as a
target space of $N=4$ sigma model \cite{BS}. The theory contains $U(1)^{r}$
gauge symmetry and $r+2$ hypermultiplets with a matrix charge $Q_{i}^{a},\;\;  a=1,\ldots,r; \quad  i=1,\ldots,r+2$.
Up to some details, this matrix can \ be identified with the Cartan matrices of the 
$ADE$ Lie algebras of rank $r$. In this way, the eight dimensional manifolds are solutions of the $N=4
$ D-flatness equations which are generally given by 
 
\begin{equation}
\sum_{i=1}^{r+2}Q_{i}^{a}[\phi _{i}^{\alpha }{\bar{\phi}}_{i\beta }+\phi
_{i}^{\beta }{\bar{\phi}}_{i\alpha }]=\vec{\xi}_{a}\vec{\sigma}_{\beta
}^{\alpha },\label{sigma4}
\end{equation}
where $\phi _{i}^{\alpha }$'s ($ \alpha=1,2)$ denote $r+2$ component field doublets of
hypermultiplets, $\vec{\xi}_{a}$ are    Fayet-Iliopolos (FI)  3-vector couplings rotated by $%
SU(2)$ symmetry, and $\vec{\sigma}_{\beta }^{\alpha }$ are the traceless $%
2\times 2$ Pauli matrices. Equations (\ref{sigma4}) deal with the
hypermultiplet branch and give a gauge invariant hyper-K\"{a}hler target
space. For each $U(1)$ factor of the $U(1)^{r}$ gauge group, they involve a
triplet of FI parameters. Using the $SU(2)$ R-symmetry  transformations $\phi
^{\alpha }=\varepsilon ^{\alpha \beta }\phi _{\beta },\;\overline{\phi
^{\alpha }}=\overline{\phi }_{\alpha },\;\varepsilon _{12}=\varepsilon
^{21}=1$ and replacing the Pauli matrices by their expressions, the
identities (\ref{sigma4}) can be split as follows 
\begin{eqnarray}
\sum\limits_{i=1}^{r+2}Q_{i}^{a}(|\phi _{i}^{1}|^{2}-|\phi _{i}^{2}|^{2})
&=&\xi _{a}^{3}  \nonumber  \label{extendedsigma} \\
\sum\limits_{i=1}^{r+2}Q_{i}^{a}\phi _{i}^{1}\overline{\phi }_{i}^{2} &=&\xi
_{a}^{1}+i{\xi ^{2}}_{a} \\
\sum\limits_{i=1}^{r+2}Q_{i}^{a}\phi _{i}^{2}\overline{\phi }_{i}^{1} &=&\xi
_{a}^{1}-i{\xi ^{2}}_{a}.  \nonumber
\end{eqnarray}%
As in the case of the $N=2$ realization of the K3 surface, the solution of
these constraints involves intersecting 4-cycles according to the   Dynkin
graphs of the ${ADE}$ Lie algebras, allowing us to produce a product of $SU$-type gauge groups with
bi-fundamental matter in F-theory comapctification with the ALE space
fibration. These models can be encoded in quiver diagrams similar to the  Dynkin
graphs.  In this graph, each node represents a $SU$ gauge group factor  and the link is associated  with matter.
Usually,  the matter is in the bi-fundamental representation of the gauge group. In this paper, we extend this techniques
to allow fundamental fields.  Roughly, we have  shown  that the base solution of (%
\ref{extendedsigma}) is given by intersecting $F_{0}=\mathbf{CP^{1}}\times 
\mathbf{CP^{1}}$ according to the $ADE$ Dynking geometries \cite{ABBS}. This is a particular
example of del Pezzo complex  surfaces which is needed for the decoupling limit of
the supergravity. This is in  agreement with the flux condition studied in 
\cite{MSS4}. From the obvious similarity with the $N=2$ scenario given  in 
\cite{KMV} it is not surprising to see some analogs  between  the  $ADE$ diagrams in
eight-dimensional hyper-K\"{a}hler manifolds and the ones considered  in the K\"{a}hler case. In fact, the intersection theory assigns the intersection number
to complex surfaces inside of these eight dimensional  manifolds. The self-intersection of the
zero section in the cotangent bundle of $F_{0}$ is equal to its minus Euler
characteristic, i.e. $-4$.  Consider now a lattice of compact 4-cycles
generated by $F_{0}^{i}$  and assume  that $F_{0}^{i}$ intersects $F_{0}^{i+1}$
at two points. This can be supported by the fact that each $\mathbf{CP^{1}}$
inside $F_{0}^{i}$ intersects just one $\mathbf{CP^{1}}$ in the next $%
F_{0}^{i+1}$. In this way, the intersection numbers of the $F_{0}$'s reads
\begin{eqnarray}
\left[ F_{0}^{i}\right] \cdot \left[ F_{0}^{i}\right]  &=&-4  \nonumber \\
\left[ F_{0}^{i}\right] \cdot \left[ F_{0}^{i+1}\right]  &=&2,
\end{eqnarray}%
with others vanishing. This means that $F_{0}^{i}$ does not intersect $%
F_{0}^{j}$ if $|j-i|>1$. Endowed with this intersection form, the lattice of these
compact 4-cycles can be identified with the root lattice of the $ADE$ \ Lie
\ algebras, up to a multiplication factor. More precisely, the relation
takes the form 
\[
\left[ F_{0}^{i}\right] \cdot \left[ F_{0}^{j}\right] =-2K^{ij}(ADE).
\]%

To describe the physics  on  seven-branes wrapped on such intersecting
geometries, we will use the result of the geometric engineering method used  in
type II superstrings \cite{KMV}. We take  different stacks of seven-branes.
If we wrap each stack of seven-branes on one $F_{0}$ complex surface, the
total gauge group takes the following form 
\begin{equation}
G\ =\prod_{i}SU(N_{i}),  \label{G}
\end{equation}%
where $i$ runs through the set of vertices of the Dynkin diagram of $ADE$
type. The matter fields are bi-fundamentals transforming in the $%
(N_{i},N_{j})$ representations. However, it has been shown in \cite{ABBS}
that the anomaly cancellation condition 
\begin{equation}
\sum \limits_{i}\left[ F_{0}^{i}\right] \cdot \left[ F_{0}^{j}\right]N_{i}=0
\end{equation}
 is translated into a condition on
the $\widehat{ADE}$ affine Lie algebra  on the base of local
F-theory compactification. In this way, $N_{i}$ should be given in terms of
Coxeter labels $s_{i}$ of the associated affine Lie algebra, namely 
\begin{equation}
N_{i}=s_{i}N.
\end{equation}
 where $N$ is an unspecified positive integer.

At  the end of this section, we would like to list  some results for $%
\widehat{AD}$ Dynkin geometries. The result for $\widehat{E}$ can be
obtained easily using the same method. For {$\widehat{A}_{r}$} for instance,
the  base of local F-theory  compactification  consists of $r+1$ intersecting $F_{0}=\mathbf{%
CP^{1}}\times \mathbf{CP^{1}}$ arranged as shown here 
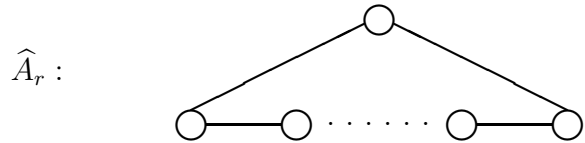
\begin{equation}
\mbox{
         \begin{picture}(20,90)(50,-20)
        \unitlength=2cm
        \thicklines
      \put(-1.2,.3){$\widehat{A}_r:$}
    \put(0,0){\circle{.2}}
     \put(.1,0){\line(1,0){.5}}
     \put(.7,0){\circle{.2}}
     \put(.9,0){$.\ .\ .\ .\ .\ .$}
     \put(1.8,0){\circle{.2}}
     \put(1.9,0){\line(1,0){.5}}
     \put(2.5,0){\circle{.2}}
     \put(0,.1){\line(2,1){1.15}}
     \put(1.25,.7){\circle{.2}}
     \put(2.5,.1){\line(-2,1){1.15}}
  \end{picture}
\nonumber}  \label{affAk}
\end{equation}%
In this graph   each node represents a  $F_{0}$ complex surface and the link is associated with the intersection.  In this case, we have $s_i=1$ and this  geometry
describes a $SU(N)^{r+1}$ quiver gauge theory with bi-fundamental field
representations.\\  In the case where the base is  a  $\widehat{D}_{r}$ Dynkin geometry 
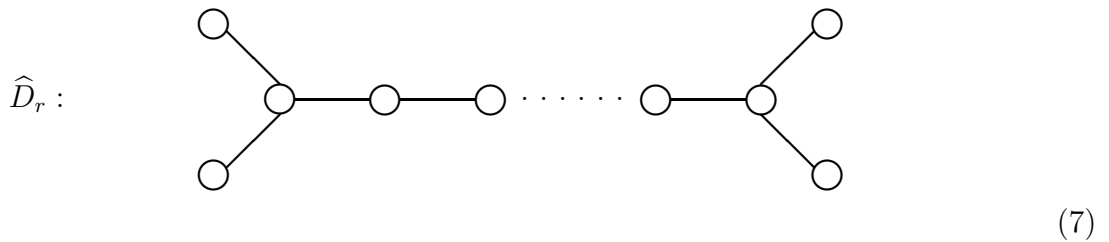
\begin{equation}
\mbox{
         \begin{picture}(20,110)(60,-50)
        \unitlength=2cm
        \thicklines
      \put(-1.8,-.05){$\widehat{D}_r:$}
    \put(0,0){\circle{.2}}
     \put(.1,0){\line(1,0){.5}}
     \put(.7,0){\circle{.2}}
     \put(.8,0){\line(1,0){.5}}
     \put(1.4,0){\circle{.2}}
     \put(1.6,0){$.\ .\ .\ .\ .\ .$}
     \put(2.5,0){\circle{.2}}
     \put(2.6,0){\line(1,0){.5}}
     \put(3.2,0){\circle{.2}}
     \put(3.2,.1){\line(1,1){.35}}
     \put(3.64,.5){\circle{.2}}
     \put(3.2,-.1){\line(1,-1){.35}}
     \put(3.64,-.5){\circle{.2}}
     \put(0,.1){\line(-1,1){.35}}
     \put(-.44,.5){\circle{.2}}
     \put(0,-.1){\line(-1,-1){.35}}
     \put(-.44,-.5){\circle{.2}}
  \end{picture}
}  \label{affDk}
\end{equation}%
the $s_{i}$ vector appearing in the gauge  group is given by $%
s_{i}=(1,1,2,\dots ,2,1,1)$. The number 1 corresponds to the monovalent nodes and 2 is associated with the remaining ones.  The gauge  group takes the form 
  $SU(N)^{4}\times SU(2N)^{r-3}$. \\ For the exceptional  base geometries,  the vector $s_i$ is as follows
\begin{eqnarray}
\widehat{E}_6 &:&s_i=(1,1,2,2,3,2,1)  \nonumber  \\
\widehat{E}_7 &:& s_i=(1,2,2,3,4,3,2,1)\\\nonumber
\widehat{E}_8 &:&s_i=(1,2,3,4,5,6,4,3,2).  \nonumber
\end{eqnarray}%
\newline
Having discussed F-theory quivers with only bi-fundamental matter, in what
follows we would like to incorporate  fundamental fields. In particular, \ we
consider a special class of hyperbolic Dynkin base geometries in the compactification of F-theory. Other possibilities will be addressed in  future work.

\section{ Fundamental Matter Fields in F-theory Quivers}

We will focus in this section on a particular base geometry producing
F-theory quivers with fundamental matter. More precisely, we will discuss
how the  fundamental fields can be obtained from intersecting
seven-branes wrapped on geometries classified by three kinds of the 
 Kac-Moody algebras. There are in general
three kinds of the $K_{ij}$'s Cartan matrices \cite{Kac,GO,ABS}. These involve
three classes defining the finite, affine and the indefinite sectors of the
Kac-Moody algebras. They are classified by the following Cartan matrices
satisfying 
\begin{eqnarray}
\sum_{j}K_{ij}^{(+)}N_{j} &=&M_{i}  \nonumber  \label{gencar} \\
\sum_{j}K_{ij}^{(0)}N_{j} &=&0 \\
\sum_{j}K_{ij}^{(-)}N_{j} &=&-M_{i},  \nonumber
\end{eqnarray}%
where $N_{j}$ and $M_{i}$ are positive integers specified later on. In these
equations, $K_{ij}^{\left( +\right) }$, $K_{ij}^{\left( 0\right) }$\ and $%
K_{ij}^{\left( -\right) }$ refer respectively to the  Cartan matrices of finite,
affine and indefinite classes. They can be put into one equation with the
following form 
\begin{equation}
\sum_{j}K_{ij}^{\left( q\right) }N_{j}\ =\ qM_{i},  \label{acc}
\end{equation}%
with $q=0,\pm 1$. In this way, $K_{ij}^{\left( q\right) }$ denotes the
generalized Cartan matrix. These equations will play a nice role in the
study of  $N=1$  F-theory quivers  in four dimensions based on Dynkin geometries.
For instance, the equation 
\begin{equation}
\sum_{j}K_{ij}^{\left( 0\right) }N_{j}=0  \label{anomalie}
\end{equation}%
has been interpreted as the anomaly cancellation condition in the affine $\widehat{ADE}$
 supersymmetric quivers without fundamental matter \cite{ABBS}. We
expect that $\sum_{j}K_{ij}^{(\pm)}N_{j}=\pm M_{i}$ should play a quite
similar role to the relation  $\sum_{j}K_{ij}^{(0)}N_{j}=0$. In fact, we will interpret the
quantities $\pm M_{i}$ as fundamental matter incorporated in F-theory
quivers encoded in Dynkin geometries. This may be seen by modifying the
anomaly cancellation condition (\ref{anomalie}) as 
\begin{equation}  \label{geanomalie}
\sum_{j}K_{ij}^{\left( q\right) }N_{j}-qM_{i}=0. 
\end{equation}%
These equations describe now the anomaly cancellation condition for the three
classes of F-theory quivers where the base  geometries are classified by
the equation (\ref{gencar}),  while $M_{i}$ denote the contributions of  fundamental
matter multiplets and where $N_{j}$ denote the colors. These numbers cannot
be arbitrary and can be fixed by (\ref{geanomalie}). It follows that one may
distinguish between two models of F-theory quivers in four dimensions according to whether or
not the model contains fundamental fields. The first model involves a quiver
theory with a gauge group as $\prod_{i}SU(s_{i}N)$ without fundamental
fields. The absence of fundamental matter  can be related the zero appearing in the second equation of (\ref{gencar}). 
 This model, which has been studied in \cite{ABBS}, has no flavor
symmetry. The second model, which we are interested in here, has an extra
flavor symmetry. In this case, we can have fundamental matter transforming
under the flavor group.

However, the general study is a highly non-trivial task as it requires
solving the above complicated matter equations (13). The general solution is
beyond the scope of the present work, though we will consider in some
details the case of a special base geometry obtained from the affine $%
\widehat{A}$ quiver. A simple situation  is to add one extra node to the
Dynkin graph of the affine $\widehat{A}$ Lie algebra. In
geometric Dynkin language, this corresponds to a particular hyperbolic
Dynkin diagram. The derivation of such a hyperbolic geometry is based on the
same philosophy one uses in the building of the  affine Dynkin diagrams from the finite ones
by adding a node. In other words, by cutting this node in  such a hyperbolic
Dynkin diagram, the resulting sub-diagram coincides with the  Dynkin
graph of the affine $\widehat{A}$ Lie algebras. The  new Dynkin graph looks like 

\begin{equation}
\mbox{
         \begin{picture}(20,90)(50,-20)
        \unitlength=2cm
        \thicklines
      \put(-1.2,.3){$\widehat{HA}_r:$}
\put(-0.7,0){\circle{.2}}
     \put(-0.6,0){\line(1,0){.5}}
    \put(0,0){\circle{.2}}
     \put(.1,0){\line(1,0){.5}}
     \put(.7,0){\circle{.2}}
     \put(.9,0){$.\ .\ .\ .\ .\ .$}
     \put(1.8,0){\circle{.2}}
     \put(1.9,0){\line(1,0){.5}}
     \put(2.5,0){\circle{.2}}
     \put(0,.1){\line(2,1){1.15}}
     \put(1.25,.7){\circle{.2}}
     \put(2.5,.1){\line(-2,1){1.15}}
  \end{picture}
}  \label{affAk}
\end{equation}%
and contains now  $r+2$ nodes. It can be viewed as an extension of the bivalent geometry
describing a linear chain of gauge groups $\prod_{j}SU$ with  only bi-fundamental
matter we have considered earlier. Recall that, in the sigma model   language the
bivalent geometry means that each vector charge takes the form 
\begin{equation}
Q_{i}=(0,\ldots, 0,1,-2,1,0,\ldots ,0).
\end{equation}%
The extra node leads to a new ingredient in F-theory quivers by
introducing the so-called trivalent geometry. In sigma model language, this
geometry contains a vector charge of the form 
\begin{equation}
Q_{i}=(-2,1,1,1,0,\ldots ,0).
\end{equation}%
Note that one can add an  extra auxiliary field  with contribution $(-1)$  in order to fulfill the Calabi-Yau condition $
\sum_i Q_{i}=0$ involved   in local $ADE$  geometries.

Roughly, the corresponding base geometry in F-theory compactification  contains   now  a central $%
F_{0}^{0}$ (which can be identified with the affine node)  with self intersection $(-4)$ intersecting three other $F_{0}$'s with contribution $(+2)$. One of them is associated with the hyperbolic 
node denoted by $F_{0}^{-1}$. Therefore, the new $F_{0}^{-1}$ satisfies the
following intersection numbers 
\begin{eqnarray}
\left[ F_{0}^{-1}\right] \cdot \left[ F_{0}^{-1}\right]  &=&-4  \nonumber \\
\left[ F_{0}^{-1}\right] \cdot \left[ F_{0}^{0}\right]  &=&2, \\
\left[ F_{0}^{-1}\right] \cdot \left[ F_{0}^{i}\right]  &=&0\quad i\neq 0. 
\nonumber
\end{eqnarray}%
In what follows, we will see that this trivalent geometry can been used to
incorporate fundamental matter in a product  of $SU$-type  gauge groups with
bi-fundamental fields. This can be seen from the fact that any Cartan
matrix $K_{ij}^{(-1)}$ of hyperbolic class can be split as 
\begin{equation}
K_{ij}^{\left( -\right) }=K_{ij}^{(0)}(\widehat{A})+\Delta _{ij}^{\left(
-\right) },  \label{ks}
\end{equation}%
where $K_{ij}^{(0)}(\widehat{A})$ denotes the Cartan matrix of the affine  $\widehat{A}$ Lie
algebras. $\Delta _{ij}$  can be viewed as matter contributions which can be interpreted
as fundamental fields.

Now let us give a concrete example of the overextended version of $\widehat{A%
}_{2}$ ($\widehat{HA}_2$). Using the trivalent geometry on the affine node, the generalized
Dynkin diagram is \begin{equation}
\mbox{
         \begin{picture}(20,90)(50,-20)
        \unitlength=2cm
        \thicklines
      \put(-1.2,.3){$\widehat{HA}_2:$}
\put(-0.7,0){\circle{.2}}
     \put(-0.6,0){\line(1,0){.5}}
    \put(0,0){\circle{.2}}
\put(0.1,0){\line(1,0){2.3}}
     \put(2.5,0){\circle{.2}}
     \put(0,.1){\line(2,1){1.15}}
     \put(1.25,.7){\circle{.2}}
     \put(2.5,.1){\line(-2,1){1.15}}
  \end{picture}
}  \label{affAk}
\end{equation}%

 and its Cartan matrix reads as 
\begin{equation}
K_{ij}^{-1}(\widehat{HA}_{2})=\left( 
\begin{array}{ccccc}
2 & -1 & 0 & 0 &  \\ 
-1 & 2 & -1 & -1 &  \\ 
0 & -1 & 2 & -1 &  \\ 
0 & -1 & -1 & 2 & 
\end{array}%
\right) , \qquad det =-3.
\end{equation}%
  For this example, the (\ref{geanomalie}) reduces to 
\begin{eqnarray}
2N_{1}-N_{2} &=&-M_{1}  \nonumber \\
-N_{1}+2N_{2}-N_{3}-N_{4} &=&-M_{2}  \nonumber \\
-N_{2}+2N_{3}-N_{4} &=&-M_{3} \\
-N_{2}-N_{3}+2N_{4} &=&-M_{4}. \nonumber
\end{eqnarray}%
The general solution of the previous equation is
\begin{eqnarray}
N_{1}&=&M_{2}+M_3+M_4       \nonumber \\
N_{2}&=&M_1+2M_2+2M_3+2M_4  \nonumber \\
N_{3}&=&M_1+2M_2+\frac{4}{3}M_3+\frac{5}{3}M_4 \\
N_{4}&=&M_1+2M_2+\frac{5}{3}M_3+\frac{4}{3}M_4. \nonumber
\end{eqnarray}%
If we put $M_3=M_4=0$, the  solution becomes 
\begin{eqnarray}
N_{1} &=&M_2, \nonumber \\
N_{2} &=& N_{3}=N_{4}=M_{1}+2M_{2}.
\end{eqnarray}%
A simple seven-brane configuration  can be given by the following  representation 
\begin{eqnarray}
N_{1} &=&0,\quad N_{2}=N_{3}=N_{4}=N  \nonumber \\
M_{1} &=&N,\quad M_{2}=M_{3}=M_{4}=0.  \nonumber
\end{eqnarray}%
This brane representation  engineers  a $SU(N)$ flavor symmetry on the hyperbolic node. In the geometric
language, this solution could be  also obtained from an extra bi-fundamental
matter associated with the hyperbolic node by assuming that the volume of $%
F_{0}^{-1}$ is very large. The Yang-Mills gauge coupling $g_{-1}$
associated with this node tends to zero. The corresponding $SU(N)$ dynamics
becomes very week, so it will be considered as a spectator flavor symmetry
group.  For $N=3$,  we get a  $SU_C(3)\times SU(3)_L\times SU(3)$  with a $SU(3)$ flavor symmetry. In this way, $SU_L(3)$ contains $SU_L(2)$; while  the 
eletric charge can have contributions from  the Cartan  sub-algebras of  $SU(3)$ and  $SU(3)_L$. Recall that $SU_C(3)\times SU(3)_L\times SU(3)$ is a maximal sub-group of the exceptional $E_6$ gauge symmetry appearing  as a possible gauge group in GUT.  In fact, the adjoint 78 representation of $E_6$  can be broken as 
\begin{equation}
78\to (8,1,1)+(1,8,1)+(1,1,8)+(3,\bar{3},3)+(3,3,\bar{3}).
 \end{equation}

\section{ Discussions}

In this work, we have  discussed F-theory quivers with bi-fundamental and
fundamental matter. Our focus has been on the building of the  base
geometry of the F-theory compactification with ALE space fibrations. The base
consists of intersecting $F_{0}=\mathbf{CP^{1}}\times \mathbf{CP^{1}}$
arranged as Dynkin graphs classified by  three kinds of  Kac-Moody algebras: ordinary finite
dimensional, affine  and indefinite  ones. We have interpreted the
equations defining these three classes of Lie algebras as the anomaly
cancelation condition of the corresponding F-theory quivers. We have
analyzed in detail hyperbolic geometries obtained from the affine $\widehat{A%
}$ geometry and we have found that it can be used to incorporate fundamental fields to a
product  of gauge groups $\prod_{j}SU$ with bi-fundamental matter. We would like to note that this analysis can be
extended by replacing each bivalent geometry of $\widehat{A}$ Dynkin graph by a 
trivalent one. This could give a chain of $SU(N)$ flavor symmetries. We also 
intend to discuss elsewhere the extension of this explicit study to the
other $\widehat{DE}$ geometries. Our approach can be adapted  to a broad
variety of geometries whose Dynkin extensions may then be obtained by adding
a node to any affine Dynkin diagram. The extension from one node to more is
straightforward. We anticipate
that this will introduce non trivial fundamental matter in F-theory
quivers based on indefinite Dynkin geometries.  This will be addressed elsewhere,  where we also intend to discuss
the implementation of polyvalent geometry  where
the bivalent and trivalent ones appear just as the leading terms of a more general case.

\textbf{Acknowledgments.}
               This work has been supported by the PCI-AECI (grant A/9335/07), Program Protars III D12/25, CICYT
(grant FPA-2006-02315) and DGIID-DGA (grant 2007-E24/2). We would like to thank E.H. Saidi for collaborations and  discussions on related subjects. AB  would
like to thank Departamento de Fisica Teorica of  Zaragoza University for kind hospitality.

\end{document}